\documentclass[prc,aps]{revtex4}
\usepackage{epsfig}

\begin{document}
\bibliographystyle{unsrt}

\title{\bf  A quark model framework for the study of nuclear medium effects}

\author{Qiang Zhao\footnote{Email address: qiang.zhao@surrey.ac.uk},
J.S. Al-Khalili\footnote{Email address: j.al-khalili@surrey.ac.uk},
and R.C. Johnson\footnote{Email address: r.johnson@surrey.ac.uk}}
\affiliation{ Department of Physics, University of Surrey, GU2 7XH, Guildford,
United Kingdom }

\begin{abstract}

A quark-model framework for studying nuclear medium effects on nucleon
resonances is described and applied here to pion photoproduction on the
deuteron, which is the simplest composite nucleon system and serves as
a first test case. Pion photoproduction on nuclei is discussed within a
chiral constituent quark model in which the quark degrees of freedom
are explicitly introduced through an effective chiral Lagrangian for
the quark-pseudoscalar-meson coupling. The advantage of this model is that a
complete set of nucleon resonances can be systematically included with
a limited number of parameters.  Also, the systematic description of
the nucleon and its resonances at quark level allows us to
self-consistently relate the nuclear medium's influence on the baryon
properties to the intrinsic dynamic aspects of the baryons.  As the
simplest composite nucleus, the deuteron represents the first
application of this effective theory for meson photoproduction on light
nuclei.  The influence of the medium on the transition operators for a
free nucleon is investigated in the Delta resonance region. No evidence
is found for a change of the Delta properties in the pion
photoproduction reaction on the deuteron since the nuclear medium here
involves just one other nucleon and the low binding energy implies low
nuclear density. However, we show that the reaction mechanism is in
principle sensitive to changes of Delta properties that would be
produced by the denser nuclear medium of heavier nuclei through the
modification of the quark model parameters.

\end{abstract}

\maketitle


\section{Introduction}

The question of the influence of the nuclear medium on the properties
of the nucleon and its resonances has attracted a lot of attention
recently both in experiment and theory.  It has also been one of the
challenges to understanding strong QCD~\cite{key-issue,k-s}
where the connection between the nucleon internal degrees of freedom
and nuclear medium effects has been a key issue.  Historically, nucleon
resonances have been among the richest sources of information on low
energy QCD phenomena.  In particular, nucleon resonance excitations by
electromagnetic probes and their decays via the strong interaction have
been an ideal method for such a study. In experiment, the advent of
high intensity photon beams at JLab, ESRF, ELSA, and MAMI has greatly
improved the data for pion production on single nucleons at the
resonance regions over the last decade.  These precise measurements
significantly improved our knowledge about nucleon resonances and
further constrained theoretical
phenomenologies~\cite{walker-69,drechsel-99,hanstein-98,aznauryan-99,crawford-nstar,davidson-91,nozawa-90-1,nozawa-90-2,sato-lee,garcilazo-93,zjli-93,arndt-96,li-97,zhao-al-li-w}.
However, our knowledge of resonance properties in nuclear
reactions~\cite{steenhoven,metag-nstar2001,rambo-99} is still very
limited, although great strides have been made by several groups to tackle
this issue with various
prescriptions~\cite{oset,korpa-lutz,lehr-mosel}.  Because of the
complexity of the strong interaction in nuclear reactions, many
phenomena may contribute, making the theoretical analyses difficult.
For instance, recent experimental work~\cite{kruche-01} on $\gamma + d\to
\pi^0 + X$ and $\gamma + \mbox{Ca}\to \pi^0 + X$ suggested a
controversial role for the $D_{13}(1520)$ by an empirical property
change of the resonance~\cite{drechsel-99-b,lehr-mosel}.  It is
essential for any theoretical model to introduce as few parameters as
possible. This is our motivation for providing a quark-level
prescription as a first step in this direction.

We propose a quark model approach
for meson photoproduction on light nuclei, based on the recent
success of a chiral effective theory for the pseudoscalar meson ($\pi$,
$\eta$, $K$) photo- and 
electroproduction~\cite{li-97,zhao-al-li-w,li-saghai-98,saghai-li-01,zhao-saghai-li-02}.
Starting with an effective chiral Lagrangian for
quark-pseudoscalar-meson coupling, this framework would in principle
permit a self-consistent inclusion 
of all the {\it s}- and {\it u}-channel resonances.  
This is a sensible approach to the
investigation of the effects of the nuclear medium on the resonances
since we believe that such effects should have an overall influence on
all the active baryons, i.e.  both the nucleon and its resonances.

In this work, we will re-visit $\gamma d\to \pi^0 d$.  The deuteron, as
the simplest composite nucleon system, provides the simplest example
for which the effective theory can be directly applied.  We will then
explore the response of baryons (described by the quark model) to the
nuclear medium, and establish the relation between nuclear medium
effects and baryon intrinsic dynamics. Of course while we acknowledge
that medium effects are hardly likely to have a significant effect for
the case of photoproduction on the deuteron, the simplicity of the
system in this preliminary work allows for a systematic study of the way
such medium effects will be manifest in heavier systems, the subject of
future work.

It will be helpful to briefly review the theoretical approaches
to $\gamma d\to \pi^0 d$ available in the literature.
The widely applied isobaric model has been successful in giving
a reasonable description of the pion cross sections
(see e.g. Ref.~\cite{blomqvist-laget-77,bosted-laget-78}).
Phenomenological methods have also been carried out based on analyses of
single nucleon interactions~\cite{lazard-76,koch-77}.
A relativistic quantum hadronic dynamical approach has also
produced similar results~\cite{garcilazo-95}.
More recently, a coupled channel approach adopting the Blomqvist-Laget 
operators~\cite{blomqvist-laget-77} was developed~\cite{k-t-b}.
In general, these approaches 
have reproduced the experimental data reasonably well in the Delta resonance 
regions due to the dominance of the Delta magnetic dipole.

A key question in nuclear reactions is the interplay of ambiguities
from the reaction mechanism and short-distance components of the
nuclear wavefunctions.  Promisingly, for the two nucleon system, such a
problem can be clarified due to the sensitivities of
transition mechanisms to certain kinematics.  This feature 
simplifies the problem and highlights the
physics of interest.  Here, we summarize several points which will
support the simplifications, and enable us to derive novel information
about the nucleon and its resonances 
in $\gamma d\to \pi^0 d$. 
Firstly, concerning the reaction mechanism, 
we neglect the rescattering contributions at the Delta resonance region
due to the dominance of the direct scattering process above the production
threshold.
This is consistent with the findings of other
models~\cite{bosted-laget-78,k-t-b}. Secondly, concerning the nuclear
wavefunction, we are interested in the role played by the
short-distance components and their effects on the cross section.  We
will show that, in the $\Delta$ resonance region, the deuteron
$D$-state generally gives corrections to the backward angle cross
sections, while the $S$-state dominates at forward angles. We will then
investigate the impact of the nuclear medium effects on both of the
kinematical regions.

\section{The model for pion photoproduction}

Let us start with an effective Lagrangian for the quark-Goldstone-boson 
interaction~\cite{manohar-georgi-84}
\begin{equation}\label{lagrangian}
{\cal L}_{eff}=\overline{\psi}[\gamma_\mu(i\partial^\mu+V^\mu 
+\gamma_5 A^\mu)-m]\psi +\cdot\cdot\cdot, 
\end{equation}
where $V^\mu$ and $A^\mu$ denote the vector and axial currents which 
have the following expressions:
\begin{eqnarray}
V_\mu &=&\frac 12 (\xi^\dag\partial_\mu\xi 
+ \xi\partial_\mu\xi^\dag),\nonumber\\
A_\mu & =& i\frac 12 (\xi^\dag\partial_\mu\xi
- \xi\partial_\mu\xi^\dag), 
\end{eqnarray}
and the chiral transformation is,
\begin{equation}\label{chiral}
\xi=e^{i\phi_m/f_m}, 
\end{equation}
where $f_m$ is the decay constant
of the meson $\phi_m$, which is the Goldstone boson field.  
In the SU(3) symmetry limit, 
the quark field $\psi$ is 
\begin{equation}  
\psi =\left( \begin{array}{c}  
\psi (u)\\ \psi (d) \\ \psi (s) \end{array} \right ),   
\end{equation}  
and the meson field $\phi_m$ is a 3$\otimes$3 matrix:
\begin{equation}  
\phi_m=\left( \begin{array}{ccc}  
\frac{1}{\sqrt{2}}\pi^{0}+\frac{1}{\sqrt{6}}\eta & \pi^{+} & K^{+}\\  
\pi^{-} & -\frac{1}{\sqrt{2}}\pi^{0}+\frac{1}{\sqrt{6}}\eta & K^{0}\\  
K^{-} & \overline{K}^{0}  &-\sqrt{\frac 23}\eta
\end{array} \right ) .
\end{equation}  
Thus, the Lagrangian in Eq.~(\ref{lagrangian})
is invariant under the chiral transformation.

Expanding the field $\xi$ in Eq.~(\ref{chiral})
in terms of $\phi_m$, i.e. 
$\xi=1+i\phi_m/f_m+\cdot\cdot\cdot$, 
we obtain the standard quark-meson
pseudovector coupling at tree level~\cite{li-97,zhao-al-li-w}:
\begin{equation}\label{coupling}
H_m=\frac{1}{f_m}\overline{\psi}\gamma_\mu
\gamma_5\partial^\mu\phi_m \psi\ ,
\end{equation}
where the Goldstone boson field $\phi_m$ is now presented by the 
SU(3) octet mesons ($\pi$, $\eta$ and $K$).

The quark-photon electromagnetic coupling is
\begin{equation}
H_{em}=-e_q\overline{\psi}\gamma_\mu A^\mu({\bf k}, {\bf r})\psi,
\end{equation}
where the photon has three momentum ${\bf k}$, and the constituent
quark carries a charge $e_q$.

This approach naturally places a constraint on the relative 
phases of those octet mesons in the SU(3) flavor symmetry limit, 
and relates the quark operators to the hadronic ones via 
explicit quark model wavefunctions for the baryons states.
In this sense, the quark model transition amplitudes can be compared
with hadronic ones, from which information about the baryon structure
can be learned.

A systematic investigation of the four channels of pion photoproduction
on a bare nucleon was carried out in~\cite{zhao-al-li-w}, where
it was shown that, up to 500 MeV photon energies, 
all the available observables could be reasonably accounted
for with only one parameter for the Delta resonance coupling.
Great effort was devoted in that work to restrict
the number of parameters by taking advantage of the quark model symmetry
in order that a complete set of resonances could be systematically
taken into account. Overall quantitative agreement of the 
model with experimental data can be seen in~\cite{zhao-al-li-w}.
In the present study, however, due to the fast fall-off
of the deuteron momentum space wavefunction at large momenta
and the photon energies considered ($E_\gamma\sim 300$ MeV),
contributions from resonances in the second resonance regions,
($P_{11}(1440)$, $S_{11}(1535)$, $D_{13}(1520)$,
$S_{31}(1620)$, $D_{33}(1700)$, $P_{33}(1600)$,
$P_{13}(1720)$, $F_{15}(1680)$~\cite{PDG2000}, etc) are generally negligible.
In the first resonance region, where
the Delta and Born term are dominant, it is a good approximation
to include only these two contributions
in the direct scattering process.

\section{Photoproduction on the deuteron}

In the $\gamma$-$d$ c.m. system
the unpolarized cross section for $\gamma d\to \pi^0 d$ can be
written as
\begin{equation}
\frac{d\sigma}{d t}=\frac{1}{64\pi s}\frac{1}{|{\bf k}|^2}
\frac{1}{2L_d+1}\sum_{m_i, m_f}|{\cal M}_{fi}(m_i,m_f,m_\gamma=+1)|^2,
\end{equation}
where ${\bf k}$
is the photon momentum defining the $z$-axis, and
 $s$ is the squared total c.m. energy; $L_d=1$ is the deuteron total spin,
 while $m_i$, $m_f$, and $m_\gamma$ denote the spin projections
 for the initial, final state deuterons, and the incoming photon, respectively.
The amplitudes with $m_\gamma=-1$ and $+1$ are not independent
due to parity conservation.

\begin{figure}
\begin{center}
\epsfig{file=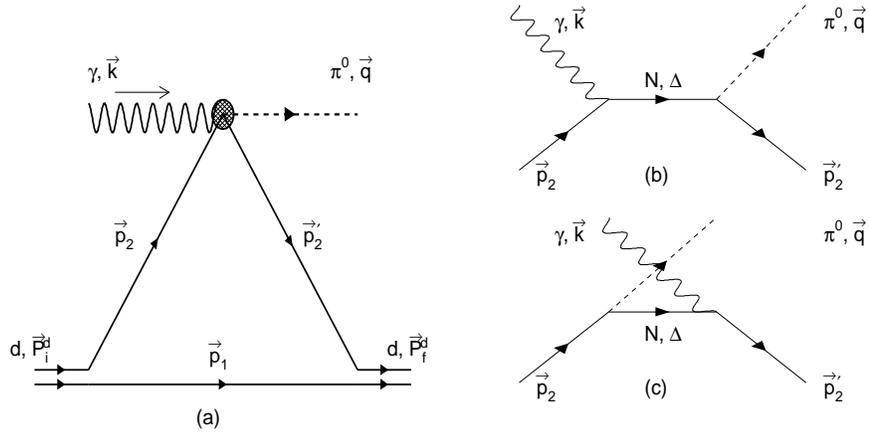,width=14cm,height=8cm}
\end{center}
\caption{Schematic diagrams for: (a) the direct scattering
process in $\gamma d\to \pi^0 d$; (b) the {\it s}-channel process
for the pion production; and (c) the {\it u}-channel process
for the pion production. In (b) and (c), the photon and pion 
will couple to the constituent quarks of the baryons as illustrated
in Ref.~\protect\cite{zhao-al-li-w}. }
\protect\label{fig:(1)}
\end{figure}

The invariant amplitude ${\cal M}_{fi}$ 
in the direct scattering 
(Fig.~\ref{fig:(1)})
can be expressed as
\begin{eqnarray}
\label{impulse}
{\cal M}_{fi} = \int \, d{\bf p}_1\ \Phi^{d*}_{m_f}({\bf p}_1+{\bf q}/2)
\ \langle\,  {\bf p}_2^\prime, {\bf q}\,| 
\,{\hat T}_{\gamma N\to \pi^0 N} |\, {\bf p}_2, {\bf k}, \rangle
\ \Phi^d_{m_i}({\bf p}_1+{\bf k}/2),
\end{eqnarray}
where $\Phi^d_m$ denotes the
deuteron ground state wavefunction with spin projection $m$;
${\hat T}_{\gamma N\to \pi^0 N}$ is the transition operator for the single
nucleon interaction derived within the quark model,
and is the nucleon photoproduction operator
averaged over
the isospin wavefunction of the deuteron.
The kinematical variables are defined as,
${\bf p}_2=-{\bf k}-{\bf p}_1$,
and ${\bf p}^\prime_2=-{\bf q}-{\bf p}_1$.

The deuteron wavefunction is defined as
\begin{eqnarray}
\Phi^d_m({\bf p})=\sum_{L=0,2; \Lambda,m_s} i^L
u_L(|{\bf p}|)Y_{L\Lambda}({\bf p}) \langle L\Lambda, 1m_s| 1 m\rangle 
\chi^d_{1m_s} ,
\end{eqnarray}
where $L=0$ and 2 denote the $S$ and $D$-state components and
the spin wavefunction is
\begin{equation}\label{d-spin}
\chi^d_{1m_s}=\sum_{m_1,m_2}
\langle \frac 12 m_1, \frac 12 m_2|1 m_s\rangle \chi_{\frac 12 m_1}
\chi_{\frac 12 m_2},
\end{equation}
where $\chi$ is the nucleon spin wavefunction, and $m_1$ and $m_2$
are spin projections for the spectator and active nucleon.

The well-established Paris model~\cite{lacombe-81} wavefunction for
the deuteron is adopted, which means ambiguities from the nuclear
wavefunctions are minimised in this examination. The Paris potential is
obtained from a phenomenological momentum-dependent meson exchange
model for the nucleon-nucleon ($N$-$N$) interaction 
with parameters chosen to fit $n$-$p$
scattering data for energies from the deuteron binding energy up to
several hundred MeV, thus giving a reliable description of the deuteron
wavefunction down to short $n$-$p$ separations.

To check the sensitivity of the reaction mechanism to the nuclear
structure, we also carry out the calculation with a Hulth\'en
wavefunction, for which the detailed description
and parameters are given in Ref.~\cite{ij-78}. This simpler wavefunction is
generated by the Yamaguchi-Yamaguchi separable $n$-$p$ interaction~\cite{yy} 
which has the
useful feature that the $S$ and $D$ components of the wavefunction have 
simple analytical forms in momentum space. The $n$-$p$ potential in this model
has parameters adjusted to give the experimental deuteron binding energy,
quadrupole moment, and $n$-$p$ triplet state effective range, 
and scattering length, 
all to high accuracy. However, it is not expected to be as accurate as the
Paris wavefunction at larger momenta.
Thus, since the long-distance asymptotic behavior of the deuteron
wavefunction is well-known, the contrast will further pin down the
nuclear structure effects arising from short-distance components.  
We neglect the $D$-state to $D$-state transition term in
the amplitude in Eq.(4) but retain the $S$ to $D$ transition. The
$D$-$D$ transition may become important at high momentum transfers and
show up in the backward-angle cross sections.  However, this is not the
focus in this work.

Firstly, we examine the single nucleon reaction operators in $\gamma
d\to \pi^0 d$ by applying the transition operators derived in the free
nucleon reaction to the direct scattering process (Fig.~\ref{fig:(1)}).
Note that the quark model parameters are determined
in the free nucleon reactions while the deuteron wavefunction used is
from independent studies. Our  calculations are therefore
parameter-free predictions of the effective theory in the quark model
framework.  
It should also be pointed out that since the final state pions
are treated as point-like particles, the size effects
and possible medium effects on the mesons are absorbed into the 
quark model parameters. 
This is different from isobaric prescriptions, where
the meson and baryon degrees of freedom could be influenced
by the nuclear medium separately
(see e.g. Refs.~\cite{oset,korpa-lutz,lehr-mosel}). 
In this study, 
we will present a new dynamic view of 
this phenomenon by relating the nuclear medium's influence
on the baryon properties 
to the intrinsic dynamic aspects of baryons via the quark model 
parameters.  We note that similar ideas are also proposed 
in Ref.~\cite{riska-brown}. 

\begin{figure}
\begin{center}
\epsfig{file=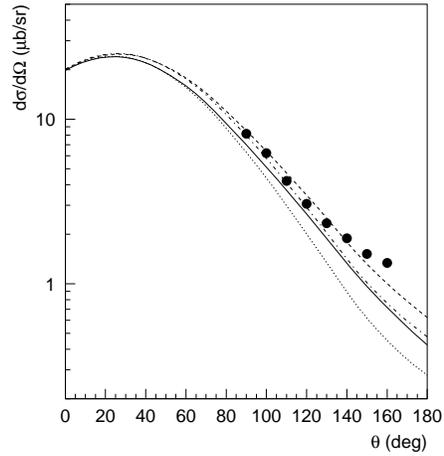, width=7cm,height=7cm}
\caption{ Angular distributions for $\gamma d\to \pi^0 d$
at $E_\gamma=300$ MeV. The solid  and dashed curve
denotes the calculations adopting the Paris and Hulth\'en wavefunctions
with both $S$ and $D$ states,
while the dotted and dot-dashed curves denote those with only
the $S$ state components, respectively. Data are from 
Ref.~\protect\cite{holtey}.
}
\protect\label{fig:(2)}
\end{center}
\end{figure}

In Fig.~\ref{fig:(2)}, the results for the differential cross sections 
at $E_\gamma=300$ MeV with both wavefunctions 
are presented.
As shown by the solid (Paris wavefunction)
and dashed curve (Hulth\'en wavefunction) in Fig.~\ref{fig:(2)}, these two
wavefunctions produce the same cross sections at forward angles due to
the same asymptotic behavior, while discrepancies arise at large angles
due to differences between these two models at short distances.  
Although the experimental data~\cite{holtey} are still sparse,
the cross 
sections using both wavefunctions are in good agreement with the data
over a wide angle region.

We also present the results without the $D$-state contributions
(the dotted and dot-dashed curve stand for the $S$-state
Paris and Hulth\'en wavefunctions, respectively),
where the large-angle cross section differences 
highlight the $D$-state contributions.

The above result is a direct examination of the 
quark model treatment for the elementary process, which deserves
more careful consideration.
Since nucleons in the deuteron are weakly bound, 
one would expect a weak effect from the nuclear medium.
The above result indeed
suggests that the baryons behave as they do 
in the single nucleon reactions. 
Therefore, in principle no parameter
change to the baryon properties is needed here. This conclusion 
is consistent with other previous 
studies~\cite{blomqvist-laget-77,bosted-laget-78,lazard-76,koch-77,garcilazo-95,k-t-b}. 
However, where our approach has an advantage is that it also allows us
to examine the impact of the nuclear medium more systematically.  We
consider a mechanism that changes the properties of the Delta resonance
due to the change of quark model parameters, e.g.  quark potential
strength within the baryons.

\section{Mechanism for medium modification}

The quark model potential strength  $\alpha_h$ and constituent 
quark mass $m_q$ 
are explicitly related to the mass 
of the baryons
in the nonrelativistic constituent quark model (NRCQM), where the nucleon
and Delta resonance belong to the same representation~\cite{close}.
A simple mass relation from the quark hyperfine interaction
will allow us to make an empirical connection:
\begin{eqnarray}
M_N(\alpha_h, m_q) &=& M_0(\alpha_h, m_q) 
-\frac 12 \delta_q(\alpha_h, m_q) \ , \nonumber\\
M_\Delta(\alpha_h, m_q) &=& M_0(\alpha_h, m_q) 
+\frac 12 \delta_q(\alpha_h, m_q) \ ,
\end{eqnarray}
where $M_0$ is the degenerate mass between the nucleon and Delta 
before the hyperfine splitting, 
and $M_N$ and $M_\Delta$ are 
the physical masses of the nucleon and Delta.
The quantity $\delta_q$ is
determined by the contact coupling:
\begin{equation}
\delta_q=\frac{4\alpha_s \alpha_h^3}{3\sqrt{2}\pi m_q^2} \ ,
\end{equation}
where $\alpha_s$ is the electromagnetic coupling. 
It is reasonable to assume, for the case of the deuteron,
that the nucleon mass $M_N$ is fixed for any set of parameters.
We can then express the mass formula as:
\begin{equation}
\label{delta-mass}
\bar{M}_\Delta(\bar{\alpha}_{h}, m_q)
=\left(\frac{\bar{\delta}_{q}}{\delta_{q}}\right)
[M_\Delta(\alpha_{h}, m_q)-M_N]+M_N \ ,
\end{equation}
where the ``barred" quantities refer to the values
given by the medium-modified parameters.

The Delta's strong decay width for $\Delta\to \pi^0 N$ can be also related 
to the quark model parameters:
\begin{eqnarray}
\label{delta-total-width}
\Gamma_\Delta &=&\left(\frac{g_{\pi N\Delta}^2}{4\pi}\right)
\frac{|{\bf q}|^3(E_f+M_N)}{8 M_\Delta M_N^2}
\left(\frac{\omega_m}{E_f+M_N}+1\right)^2 \nonumber\\
& &\times e^{-{\bf q}^2/3\alpha_h^2} \ ,
\end{eqnarray}
where $E_f=(M_N^2+|{\bf q}|^2)^{1/2}$ and 
$\omega_m=(m_\pi^2 +|{\bf q}|^2)^{1/2}$ are the nucleon and pion energies
in the Delta c.m. system. These variables are determined by the Delta mass, 
which hence are related to Eq.~(\ref{delta-mass}). Also,
the change of the potential strength will influence the form factors 
for baryons, i.e. both nucleon and Delta. 
Therefore, even though we fix the mass of the nucleon, its size
may change due to the change of quark potential strength in the nuclear medium.
For the Delta, we can simply express
$\Gamma_\Delta\equiv\Gamma_\Delta(\alpha_h, m_q)$.
The $\pi N\Delta$ coupling $g_{\pi N\Delta}\equiv g_{\pi NN}C_{\pi N\Delta}$
is determined in the single free nucleon reaction~\cite{zhao-al-li-w}, 
where $C_{\pi N\Delta}$ is the strength given 
by the experimental data.

The simple relations from the static properties of baryons
introduces new aspects in the reaction mechanism.
If the medium influence leads to a significant change 
to $\alpha_h$ and $m_q$, from Eq.~(\ref{delta-mass}) 
and (\ref{delta-total-width}), 
the mass shift and the width change of the Delta will be correlated.
As a sensitivity test, we assume  the quark mass is fixed, and find 
that 
a change of the quark potential strength from $\alpha_h=330.31$ MeV 
(determined in the free nucleon reaction~\cite{zhao-al-li-w}) to
$\bar{\alpha}_{h}=342$ MeV 
results in 
a mass increase of about 36 MeV as suggested by other calculations 
and total width change from $\Gamma_{\Delta}=120$ MeV  
to $\bar{\Gamma}_{\Delta}=156.7$ MeV.

\begin{figure}
\begin{center}
\epsfig{file=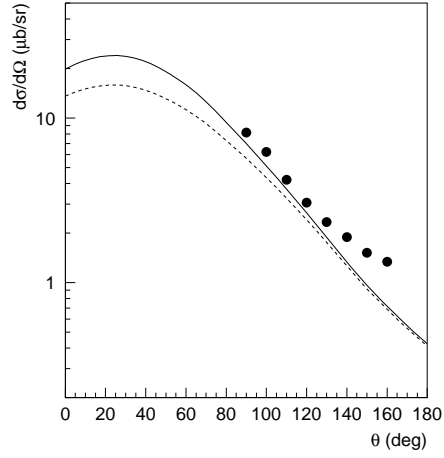, width=7cm,height=7cm}
\caption{ Estimation of effects from the Delta property change. 
The solid curve is the result of the unchanged-parameter calculation 
(the same as the solid one in Fig.~\protect\ref{fig:(2)}).
The dashed curve denotes the
results for a changed Delta with a flattened width and larger mass.
Data are the same as in Fig.~\protect\ref{fig:(2)}.
}
\protect\label{fig:(3)}
\end{center}
\end{figure}

In Fig.~\ref{fig:(3)}, we present the calculated
results (dashed curve) for the differential cross sections
with such a property change to the baryons. 
By comparison with 
the unchanged-parameter calculation (solid curve),
we see that the cross section is very sensitive to the 
quark model parameters at forward angles. 
This can be understood as follows. The overall shape of the cross section
is governed by the behaviour of the deuteron wavefunctions, 
which ensure that the integrand in
Eq.~(\ref{impulse}) is only non-negligible 
when the arguments of both wavefunctions
are small. 
This is when ${\bf p}_1$ is of the same order as both ${\bf k}/2$
and ${\bf q}/2$, and when ${\bf k}$ and ${\bf q}$ are parallel (forward
scattering). 
In the c.m. frame we also have ${\bf p}_1 + {\bf p}_2 = -{\bf k}$ and hence 
at forward angles where ${\bf p}_1$ is large, 
${\bf p}_2$, the momentum of the active nucleon, is
at its smallest; at larger scattering angles it carries away more 
momentum from its interaction with the photon.
Since the photon energy is chosen to be close to 
the Delta threshold in the photon-active nucleon lab system, 
the forward angles correspond to the kinematics where
the Delta will contribute 
significantly to the cross section, and medium effects on
the Delta will also show up most here.
At backward angles, most of the momenta carried by the deuteron
will be on the active nucleon, resulting in a 
shift of the photon-active-nucleon c.m. energy far away from 
the Delta mass position leading to the small effects
from the parameter change.

The fact that the overall shape of the cross section
is governed by the behaviour of the deuteron wavefunctions,
also accounts for the negligible effects from the 
energy conservation breaking at the initial photon-nucleon and final  
meson-nucleon interaction vertices in this reaction. 
For denser nuclear medium, 
such effects may become important due to larger contributions
from short-range components of nuclear wavefunctions.
This could be a possible mechanism 
by which the nuclear medium can 
change the quark model parameters.

\begin{figure}
\begin{center}
\epsfig{file=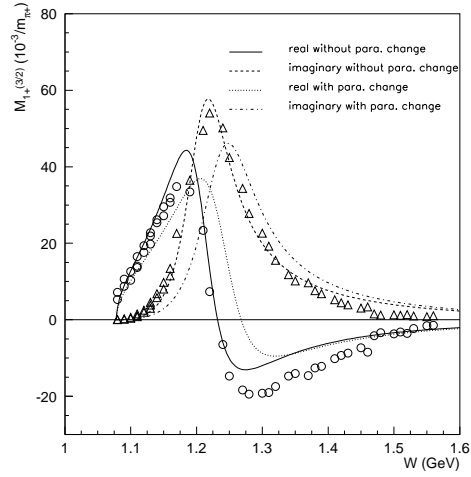, width=7cm,height=7cm}
\caption{ The Delta magnetic dipole plotted in term of the photon 
energy in the $\gamma N$ lab system. 
Data are from SAID analyses~\protect\cite{said}.
}
\protect\label{fig:(4)}
\end{center}
\end{figure}

In Fig.~\ref{fig:(4)}, the medium influence 
on the Delta magnetic dipole is compared
with the free nucleon case in the photon-nucleon c.m. frame.
The largest effects are from the region where the 
Delta is excited close to its mass shell. 
As discussed above, for a photon energy $E_\gamma=300$ MeV,
the close-to-threshold region corresponds to 
the forward scatterings, with the off-shell region to 
the backward angles. When the Delta goes far off-shell,
the parameter change effects become negligible.
Thus, one would expect that the forward angle 
scattering is the ideal place for the investigation of the Delta 
property changes due to the medium.

It is noted that the major contributions 
to the small angle cross section come from deuteron configurations 
in which the neutron-proton ($n$-$p$) separation is larger than 
the range of strong interactions. For these 
configurations any medium modifications to $\alpha_h$ will be 
negligible. Figure~\ref{fig:(3)} should therefore 
only be used as an indication of the 
relative sensitivity of the pion production process at different momentum 
transfers to a `fixed' density-independent change to $\alpha_h$. 
Any density dependent mechanism which produced a  change 
at small $n$-$p$ separations (and 
large momentum transfer) of the magnitude used in our sensitivity test would be 
expected to be associated with much smaller changes 
at large $n$-$p$ separations 
and small momentum transfers.

\section{Summary}

We have studied the photoproduction of $\pi^0$ on the
deuteron at the energy of the Delta excitation in a quark model
framework with an effective chiral Lagrangian.  The advantage of this
approach has been that the nucleon and its resonances were introduced
into the nuclear reaction with only a limited number of
parameters~\cite{zhao-al-li-w}.  In this approach we propose that the
nuclear medium effects on the baryon properties can be recognized
through the change of quark model parameters for baryons.  In
particular, the nuclear medium may influence the quark potential
strengths of a baryon state, and further lead to changes to its mass
and width.  It  thus naturally relates the resonance property changes
to their intrinsic dynamical aspects.  Mechanisms by which the
nuclear medium changes the quark model parameters need to be addressed
by investigating reactions on heavier and denser nuclei.

We expect that a systematic investigation of reactions on heavier nuclei will
establish the medium-density-dependence of the quark model
parameters.  It is also important to extend this approach to the second
resonance region, where higher resonances such as $P_{11}(1440)$,
$D_{13}(1520)$, $S_{11}(1535)$, etc., can be investigated.
Consequent work will be reported elsewhere.

Authors thank J.A. Tostevin for providing the Paris wavefunction and 
useful discussions. Useful comments from Z.P. Li, C. Bennhold and 
R.L. Workman on an early version of this paper are greatly appreciated. 
Useful discussions with E. Oset and T.T.S. Kuo are gratefully acknowledged.
Financial supports of the U.K. Engineering and Physical
Sciences Research Council (Grant No. GR/R78633/01 and GR/M82141) 
are gratefully
acknowledged.



\end{document}